\documentclass{svmult}
\usepackage{epsfig,graphicx} 
\usepackage[bottom]{footmisc}
\usepackage{natbib,url}      
\bibpunct{(}{)}{;}{a}{}{,}   
\def\cite#1{\citealp{#1}}    
\def\authorindex#1{}  
\def\figspath{.}  

\begin{document}\newcount\preprintheader\preprintheader=1



\def\thisvolume{these proceedings}

\def\aj{{AJ}}			
\def\araa{{ARA\&A}}		
\def\apj{{ApJ}}			
\def\apjl{{ApJ}}		
\def\apjs{{ApJS}}		
\def\ao{{Appl.\ Optics}} 
\def\apss{{Ap\&SS}}		
\def\aap{{A\&A}}		
\def\aapr{{A\&A~Rev.}}		
\def\aaps{{A\&AS}}		
\def\an{{Astron.\ Nachrichten}}
\def\aspcs{{ASP Conf.\ Ser.}}
\def\assp{{Astrophys.\ \& Space Sci.\ Procs., Springer, Heidelberg}}
\def\azh{{AZh}}			
\def\baas{{BAAS}}		
\def\jrasc{{JRASC}}	
\def\memras{{MmRAS}}		
\def\mnras{{MNRAS}}
\def\nat{{Nat}}		
\def\pra{{Phys.\ Rev.\ A}} 
\def\prb{{Phys.\ Rev.\ B}}		
\def\prc{{Phys.\ Rev.\ C}}		
\def\prd{{Phys.\ Rev.\ D}}		
\def\prl{{Phys.\ Rev.\ Lett.}} 
\def\pasp{{PASP}}
\def\pasj{{PASJ}}		
\def\qjras{{QJRAS}}
\def\science{{Sci}}		
\def\skytel{{S\&T}}		
\def\solphys{{Solar\ Phys.}} 
\def\sovast{{Soviet\ Ast.}}  
\def\ssr{{Space\ Sci.\ Rev.}}
\def\svassp{{Astrophys.\ Space Sci.\ Procs., Springer, Heidelberg}}
\def\zap{{ZAp}}			
\let\astap=\aap
\let\apjlett=\apjl
\let\apjsupp=\apjs
\def\grl{{Geophys.\ Res.\ Lett.}}  
\def\jgr{{J. Geophys.\ Res.}} 

\def\ion#1#2{{\rm #1}\,{\uppercase{#2}}}  
\def\deg{\hbox{$^\circ$}}
\def\sun{\hbox{$\odot$}}
\def\earth{\hbox{$\oplus$}}
\def\la{\mathrel{\hbox{\rlap{\hbox{\lower4pt\hbox{$\sim$}}}\hbox{$<$}}}}
\def\ga{\mathrel{\hbox{\rlap{\hbox{\lower4pt\hbox{$\sim$}}}\hbox{$>$}}}}
\def\sq{\hbox{\rlap{$\sqcap$}$\sqcup$}}
\def\arcmin{\hbox{$^\prime$}}
\def\arcsec{\hbox{$^{\prime\prime}$}}
\def\fd{\hbox{$.\!\!^{\rm d}$}}
\def\fh{\hbox{$.\!\!^{\rm h}$}}
\def\fm{\hbox{$.\!\!^{\rm m}$}}
\def\fs{\hbox{$.\!\!^{\rm s}$}}
\def\fdg{\hbox{$.\!\!^\circ$}}
\def\farcm{\hbox{$.\mkern-4mu^\prime$}}
\def\farcs{\hbox{$.\!\!^{\prime\prime}$}}
\def\fp{\hbox{$.\!\!^{\scriptscriptstyle\rm p}$}}
\def\micron{\hbox{$\mu$m}}
\def\onehalf{\hbox{$\,^1\!/_2$}}	
\def\onethird{\hbox{$\,^1\!/_3$}}
\def\twothirds{\hbox{$\,^2\!/_3$}}
\def\onequarter{\hbox{$\,^1\!/_4$}}
\def\threequarters{\hbox{$\,^3\!/_4$}}
\def\ubv{\hbox{$U\!BV$}}		
\def\ubvr{\hbox{$U\!BV\!R$}}		
\def\ubvri{\hbox{$U\!BV\!RI$}}		
\def\ubvrij{\hbox{$U\!BV\!RI\!J$}}		
\def\ubvrijh{\hbox{$U\!BV\!RI\!J\!H$}}		
\def\ubvrijhk{\hbox{$U\!BV\!RI\!J\!H\!K$}}		
\def\ub{\hbox{$U\!-\!B$}}		
\def\bv{\hbox{$B\!-\!V$}}		
\def\vr{\hbox{$V\!-\!R$}}		
\def\ur{\hbox{$U\!-\!R$}}


\def\labelitemi{{\bf --}}  

\def\rmit#1{{\it #1}}              
\def\rmit#1{{\rm #1}}              
\def\etal{\rmit{et al.}}           
\def\etc{\rmit{etc.}}           
\def\ie{\rmit{i.e.,}}              
\def\eg{\rmit{e.g.,}}              
\def\cf{cf.}                       
\def\viz{\rmit{viz.}}
\def\vs{\rmit{vs.}}

\def\rot{\hbox{\rm rot}}
\def\div{\hbox{\rm div}}
\def\lesssim{\mathrel{\hbox{\rlap{\hbox{\lower4pt\hbox{$\sim$}}}\hbox{$<$}}}}
\def\gtrsim{\mathrel{\hbox{\rlap{\hbox{\lower4pt\hbox{$\sim$}}}\hbox{$>$}}}}
\def\dif{\: {\rm d}}                       
\def\ep{\:{\rm e}^}                        
\def\dash{\hbox{$\,-\,$}}                  
\def\is{\!=\!}                             

\def\starname#1#2{${#1}$\,{\rm {#2}}}  
\def\Teff{\hbox{$T_{\rm eff}$}}   

\def\kms{\hbox{km$\;$s$^{-1}$}}
\def\ms{\hbox{m$\;$s$^{-1}$}}
\def\Mxcm{\hbox{Mx\,cm$^{-2}$}}    

\def\Bapp{\hbox{$B_{\rm app}$}}    

\def\komega{($k, \omega$)}                 
\def\kf{($k_h,f$)}                         
\def\VminI{\hbox{$V\!\!-\!\!I$}}           
\def\IminI{\hbox{$I\!\!-\!\!I$}}           
\def\VminV{\hbox{$V\!\!-\!\!V$}}           
\def\Xt{\hbox{$X\!\!-\!t$}}                

\def\level #1 #2#3#4{$#1 \: ^{#2} \mbox{#3} ^{#4}$}   

\def\specchar#1{\uppercase{#1}}    
\def\AlI{\mbox{Al\,\specchar{i}}}  
\def\BI{\mbox{B\,\specchar{i}}} 
\def\BII{\mbox{B\,\specchar{ii}}}  
\def\BaI{\mbox{Ba\,\specchar{i}}}  
\def\BaII{\mbox{Ba\,\specchar{ii}}} 
\def\CI{\mbox{C\,\specchar{i}}} 
\def\CII{\mbox{C\,\specchar{ii}}} 
\def\CIII{\mbox{C\,\specchar{iii}}} 
\def\CIV{\mbox{C\,\specchar{iv}}} 
\def\CaI{\mbox{Ca\,\specchar{i}}} 
\def\CaII{\mbox{Ca\,\specchar{ii}}} 
\def\CaIII{\mbox{Ca\,\specchar{iii}}} 
\def\CoI{\mbox{Co\,\specchar{i}}} 
\def\CrI{\mbox{Cr\,\specchar{i}}} 
\def\CriI{\mbox{Cr\,\specchar{ii}}} 
\def\CsI{\mbox{Cs\,\specchar{i}}} 
\def\CsII{\mbox{Cs\,\specchar{ii}}} 
\def\CuI{\mbox{Cu\,\specchar{i}}} 
\def\FeI{\mbox{Fe\,\specchar{i}}} 
\def\FeII{\mbox{Fe\,\specchar{ii}}} 
\def\FeIX{\mbox{Fe\,\specchar{ix}}}
\def\FeX{\mbox{Fe\,\specchar{x}}}
\def\FeXVI{\mbox{Fe\,\specchar{xvi}}}
\def\FrI{\mbox{Fr\,\specchar{i}}}
\def\HI{\mbox{H\,\specchar{i}}} 
\def\HII{\mbox{H\,\specchar{ii}}} 
\def\Hmin{\hbox{\rmH$^{^{_{\scriptstyle -}}}$}}      
\def\Hemin{\hbox{{\rm He}$^{^{_{\scriptstyle -}}}$}} 
\def\HeI{\mbox{He\,\specchar{i}}} 
\def\HeII{\mbox{He\,\specchar{ii}}} 
\def\HeIII{\mbox{He\,\specchar{iii}}} 
\def\KI{\mbox{K\,\specchar{i}}} 
\def\KII{\mbox{K\,\specchar{ii}}} 
\def\KIII{\mbox{K\,\specchar{iii}}} 
\def\LiI{\mbox{Li\,\specchar{i}}} 
\def\LiII{\mbox{Li\,\specchar{ii}}} 
\def\LiIII{\mbox{Li\,\specchar{iii}}} 
\def\MgI{\mbox{Mg\,\specchar{i}}} 
\def\MgII{\mbox{Mg\,\specchar{ii}}} 
\def\MgIII{\mbox{Mg\,\specchar{iii}}} 
\def\MnI{\mbox{Mn\,\specchar{i}}} 
\def\NI{\mbox{N\,\specchar{i}}}
\def\NIV{\mbox{N\,\specchar{iv}}}
\def\NaI{\mbox{Na\,\specchar{i}}}
\def\NaII{\mbox{Na\,\specchar{ii}}}
\def\NaIII{\mbox{Na\,\specchar{iii}}}
\def\NeVIII{\mbox{Ne\,\specchar{viii}}} 
\def\NiI{\mbox{Ni\,\specchar{i}}} 
\def\NiII{\mbox{Ni\,\specchar{ii}}}
\def\NiIII{\mbox{Ni\,\specchar{iii}}} 
\def\OI{\mbox{O\,\specchar{i}}} 
\def\OVI{\mbox{O\,\specchar{vi}}}
\def\RbI{\mbox{Rb\,\specchar{i}}} 
\def\SII{\mbox{S\,\specchar{ii}}} 
\def\SiI{\mbox{Si\,\specchar{i}}} 
\def\SiII{\mbox{Si\,\specchar{ii}}} 
\def\SrI{\mbox{Sr\,\specchar{i}}}
\def\SrII{\mbox{Sr\,\specchar{ii}}}
\def\TiI{\mbox{Ti\,\specchar{i}}} 
\def\TiII{\mbox{Ti\,\specchar{ii}}} 
\def\TiIII{\mbox{Ti\,\specchar{iii}}} 
\def\TiIV{\mbox{Ti\,\specchar{iv}}} 
\def\VI{\mbox{V\,\specchar{i}}} 
\def\HtwoO{\mbox{H$_2$O}}        
\def\Otwo{\mbox{O$_2$}}          

\def\Halpha{\mbox{H\hspace{0.1ex}$\alpha$}} 
\def\Ha{\mbox{H\hspace{0.2ex}$\alpha$}}
\def\Hbeta{\mbox{H\hspace{0.2ex}$\beta$}}
\def\Hgamma{\mbox{H\hspace{0.2ex}$\gamma$}}
\def\Hdelta{\mbox{H\hspace{0.2ex}$\delta$}}
\def\Hepsilon{\mbox{H\hspace{0.2ex}$\epsilon$}}
\def\Hzeta{\mbox{H\hspace{0.2ex}$\zeta$}}
\def\Lyalpha{\mbox{Ly$\hspace{0.2ex}\alpha$}}
\def\Lybeta{\mbox{Ly$\hspace{0.2ex}\beta$}}
\def\Lygamma{\mbox{Ly$\hspace{0.2ex}\gamma$}}
\def\Lycont{\mbox{Ly\hspace{0.2ex}{\small cont}}}
\def\Baalpha{\mbox{Ba$\hspace{0.2ex}\alpha$}}
\def\Babeta{\mbox{Ba$\hspace{0.2ex}\beta$}}
\def\Bacont{\mbox{Ba\hspace{0.2ex}{\small cont}}}
\def\Paalpha{\mbox{Pa$\hspace{0.2ex}\alpha$}}
\def\Bralpha{\mbox{Br$\hspace{0.2ex}\alpha$}}

\def\NaD{\mbox{Na\,\specchar{i}\,D}}    
\def\NaDone{\mbox{Na\,\specchar{i}\,\,D$_1$}}
\def\NaDtwo{\mbox{Na\,\specchar{i}\,\,D$_2$}}
\def\NaID{\mbox{Na\,\specchar{i}\,\,D}}
\def\NaIDone{\mbox{Na\,\specchar{i}\,\,D$_1$}}
\def\NaIDtwo{\mbox{Na\,\specchar{i}\,\,D$_2$}}
\def\Done{\mbox{D$_1$}}
\def\Dtwo{\mbox{D$_2$}}

\def\Mgbone{\mbox{Mg\,\specchar{i}\,b$_1$}}
\def\Mgbtwo{\mbox{Mg\,\specchar{i}\,b$_2$}}
\def\Mgbthree{\mbox{Mg\,\specchar{i}\,b$_3$}}
\def\MgIb{\mbox{Mg\,\specchar{i}\,b}}
\def\MgIbone{\mbox{Mg\,\specchar{i}\,b$_1$}}
\def\MgIbtwo{\mbox{Mg\,\specchar{i}\,b$_2$}}
\def\MgIbthree{\mbox{Mg\,\specchar{i}\,b$_3$}}

\def\CaIIK{\mbox{Ca\,\specchar{ii}\,K}}       
\def\CaIIH{\mbox{Ca\,\specchar{ii}\,H}}
\def\CaIIHK{\mbox{Ca\,\specchar{ii}\,H\,\&\,K}}
\def\HK{\mbox{H\,\&\,K}}
\def\Kthree{\mbox{K$_3$}}      
\def\Hthree{\mbox{H$_3$}}
\def\Ktwo{\mbox{K$_2$}}
\def\Htwo{\mbox{H$_2$}}
\def\Kone{\mbox{K$_1$}}     
\def\Hone{\mbox{H$_1$}}     
\def\KtwoV{\mbox{K$_{2V}$}}
\def\KtwoR{\mbox{K$_{2R}$}}
\def\KoneV{\mbox{K$_{1V}$}}
\def\KoneR{\mbox{K$_{1R}$}}
\def\HtwoV{\mbox{H$_{2V}$}}
\def\HtwoR{\mbox{H$_{2R}$}}
\def\HoneV{\mbox{H$_{1V}$}}
\def\HoneR{\mbox{H$_{1R}$}}

\def\hk{\mbox{h\,\&\,k}}
\def\kthree{\mbox{k$_3$}}    
\def\hthree{\mbox{h$_3$}}
\def\ktwo{\mbox{k$_2$}}
\def\htwo{\mbox{h$_2$}}
\def\kone{\mbox{k$_1$}}     
\def\hone{\mbox{h$_1$}}     
\def\ktwoV{\mbox{k$_{2V}$}}
\def\ktwoR{\mbox{k$_{2R}$}}
\def\koneV{\mbox{k$_{1V}$}}
\def\koneR{\mbox{k$_{1R}$}}
\def\htwoV{\mbox{h$_{2V}$}}
\def\htwoR{\mbox{h$_{2R}$}}
\def\honeV{\mbox{h$_{1V}$}}
\def\honeR{\mbox{h$_{1R}$}}

\ifnum\preprintheader=1     
\makeatletter  
\def\@maketitle{\newpage
\markboth{}{}%
  {\em \footnotesize To appear in ``Magnetic Coupling between the Interior 
       and the Atmosphere of the Sun'', eds. S.~S.~Hasan and R.~J.~Rutten, 
       Astrophysics and Space Science Proceedings, Springer-Verlag, 
       Heidelberg, Berlin, 2009.}\par
 \def\lastand{\ifnum\value{@inst}=2\relax
                 \unskip{} \andname\
              \else
                 \unskip \lastandname\
              \fi}%
 \def\and{\stepcounter{@auth}\relax
          \ifnum\value{@auth}=\value{@inst}%
             \lastand
          \else
             \unskip,
          \fi}%
  \raggedright
 {\Large \bfseries\boldmath
  \pretolerance=10000
  \let\\=\newline
  \raggedright
  \hyphenpenalty \@M
  \interlinepenalty \@M
  \if@numart
     \chap@hangfrom{}
  \else
     \chap@hangfrom{\thechapter\thechapterend\hskip\betweenumberspace}
  \fi
  \ignorespaces
  \@title \par}\vskip .8cm
\if!\@subtitle!\else {\large \bfseries\boldmath
  \vskip -.65cm
  \pretolerance=10000
  \@subtitle \par}\vskip .8cm\fi
 \setbox0=\vbox{\setcounter{@auth}{1}\def\and{\stepcounter{@auth}}%
 \def\thanks##1{}\@author}%
 \global\value{@inst}=\value{@auth}%
 \global\value{auco}=\value{@auth}%
 \setcounter{@auth}{1}%
{\lineskip .5em
\noindent\ignorespaces
\@author\vskip.35cm}
 {\small\institutename\par}
 \ifdim\pagetotal>157\p@
     \vskip 11\p@
 \else
     \@tempdima=168\p@\advance\@tempdima by-\pagetotal
     \vskip\@tempdima
 \fi
}
\makeatother     
\fi


\title*{The Magnetic Field of Solar Spicules}


\author{R. Centeno \inst{1,2} 
\and
J. Trujillo Bueno\inst{2,3} 
\and 
A. Asensio Ramos\inst{2}}


\authorindex{Centeno, R.}
\authorindex{Trujillo Bueno, J.}
\authorindex{Asensio Ramos, A.}
 


\institute{High Altitude Observatory, Boulder (USA)
\and
Instituto de Astrof\'\i sica de Canarias, La Laguna (Spain)
\and
Consejo Superior de Investigaciones Cient\'\i ficas (Spain)}

\maketitle

\setcounter{footnote}{0}  

\begin{abstract} 
Determining the magnetic field of solar spicules is vital for
developing adequate models of these plasma jets, which are thought to
play a key role in the thermal, dynamic, and magnetic structure of the
chromosphere.  Here we report on magnetic spicule properties in a very
quiet region of the off-limb solar atmosphere, as inferred from new
spectropolarimetric observations in the \HeI\,10830\,\AA\
triplet. We have used a novel inversion code for Stokes profiles
caused by the joint action of atomic level polarization and the Hanle
and Zeeman effects (HAZEL) to interpret the observations. Magnetic
fields as strong as 40~G were unambiguously detected in a very
localized area of the slit, which may represent a possible lower
value of the field strength of organized network spicules.
\end{abstract}

\section{Introduction}      \label{centeno:introduction}

The first observational evidence of solar spicules came in the
drawings of Father Angelo Secchi in the late nineteenth century.  He
recorded the shape of these off-limb jet-like structures and
listed some of their properties.  Spicules can be described as rapidly
evolving chromospheric plasma jets, protruding outside the solar limb
into the corona. It is thought that they constitute an important
ingredient of the mass balance of the solar atmosphere, since they are
estimated to carry about 100 times the mass of the solar wind.  After
the spicular material is shot up (with typical apparent velocities
around 25~kms$^{-1}$) and has reached its maximum height, it returns
to the surface along the same path or a different one.
However, many spicules do not seem to retract, but rather fade away
(see \cite{depontieu2007}).  The average direction of spicules
deviates from the vertical, reaching typical heights of $6500 -
9500$~km. Observations yield densities aboout ($3\,10^{-13}$~g/cm$^3$)
and temperatures in the range $5000 - 15000$~K that seem to be
constant with height (see \cite{beckers-spicules} for an early review
of spicule properties).

All classical spicule models make use of a magnetic flux tube that
expands from the photosphere all the way up into the corona as their
main ingredient (e.g., \cite{sterling-spicules}, \cite{DePontieu}). An
injection of energy into the flux tube is required to launch the
material and to raise it up to heights of several thousand kilometers.
Although the various models seem to explain some of the observational
aspects of quiet-Sun spicules, they all fail to reproduce one or other
observed parameter. One of the key impediments is our poor knowledge
of the magnetic properties of spicules. We need more and better
observations to constrain the models, and this is what motivates the
present investigation. We want to find reliable constraints of some of
the physical aspects of spicules, focusing, in particular, on
understanding the magnetic field topology and its behaviour along the
length of the spicule.

We use new spectropolarimetric measurements of the \HeI\,10830\,\AA\
multiplet beyond the limb to infer the magnetic properties of
quiet-Sun spicules. The information was retrieved by inverting the
observed Stokes profiles (caused by atomic level polarization and the
Hanle and Zeeman effects), as was first done by
\citet{tb-spicules}. These authors inferred strengths of 10~G in
quiet-Sun spicules at an atmospheric height of 2000~km. They point
out, however, that significantly stronger fields could also be present
(as indicated by larger Stokes-$V$ signals detected during
another observing run). The possibility of magnetic fields
significantly larger than 10~G was also tentatively suggested by the
\HeI\ D$_3$ measurements of \citet{lopezariste-spicules},
although these spicules emanated from active plage. More
detailed spectropolarimetric observations of spicules in the 
\HeI~D$_3$ multiplet were carried out by \citet{ramelli-spicules}, who
found $B{\approx}10$~G in quiet Sun and $B{\approx}50$~G in more
active areas.  The presented \HeI\,10830\,\AA\ spectropolarimetric
observations of quiet-Sun spicules provide an unambiguous
demonstration that the magnetic field strength of some spicules can be
significantly large.

\section{Observations}\label{centeno-observations}

Observations were carried out with the Tenerife Infrared Polarimeter
(TIP, \cite{martinezpillet-tip}) at the German Vacuum Tower Telescope
(Tenerife, Spain) on August 17, 2008.  The TIP instrument allowed us
to measure (almost) simultaneously the full Stokes vector of the
10830\,\AA\ spectral region for all the points along the spectrograph
slit, with spectral and spatial samplings of 11 m\AA\ and 0.17\arcsec,
respectively.  Standard data reduction routines were applied to all
the data-sets, encompassing dark current and flat-field correction of
the images as well as polarization calibration and cross-talk
correction of the Stokes profiles.

We placed the slit 2\arcsec\ off, and parallel to, the visible South
limb, crossing a forest of spicules. There, we carried out several
time series with the slit at fixed distances to the visible limb.  The
seeing conditions were not optimal and the off-limb pointing rendered
the Adaptive Optics system inoperable. However, the atmospheric
conditions were very stable (no wind), 
warranting image stability with no coelostat vibrations during the
runs. Each data-set of 50-min duration was averaged in time in order
to obtain a large S/N ratio. Spectral and spatial pixel binning were
performed for the same purpose, maintaining sufficient sampling
($\approx 0.7$\arcsec).

\section{Analysis}\label{centeno-analysis}

Solar magnetic fields leave their fingerprints on the emergent
polarization patterns of spectral lines that form in the solar
atmosphere. This occurs through the Hanle and Zeeman effects.

The spectral line polarization produced by the Zeeman effect is a
consequence of the wavelength shifts between the $\pi$ and $\sigma$
components of the atomic transitions, as the energy levels split due
to the presence of a magnetic field. This splitting is normally
proportional to the magnetic field strength and the Land\'e factor of
the level. Typically, fields of 100~G or more are needed to be able to
observe the signature of the transverse Zeeman effect on the Stokes
$Q$ and $U$ profiles of a spectral line, while much weaker resolved
fields are enough to produce measurable Stokes-$V$ signals via the
longitudinal Zeeman effect. However, when the magnetic field is too
weak and/or when there are mixed magnetic polarities within the
spatio-temporal resolution element, the circular polarization produced
by the longitudinal Zeeman effect tends to be negligible.

Fortunately, even in the absence of magnetic fields, measurable
polarization signals in a spectral line occur if there are population
imbalances among the magnetic sublevels of the atom. The key mechanism
that produces this so-called atomic level polarization in the solar
atmosphere is the anisotropic illumination of the atoms.  Such
``optical pumping'' needs no magnetic field to operate and it is very
effective in generating atomic level polarization when the
depolarizing rates from elastic collisions are low. Structures such as
chromospheric spicules are subject to the center-to-limb (CLV)
variation of the photospheric illumination, receiving more radiation
from the plasma that is directly underneath them than from the sides.

The Hanle effect is the modification of the atomic level polarization
due to the presence of a magnetic field inclined with respect to the
axis of symmetry of the radiation field. It is sensitive to weaker
magnetic fields than those needed to induce a measurable Zeeman
polarization signal and it does not tend to cancel out when mixed
polarities are present (see \cite{tb-quantumspectropol}).  The
observational signatures of the Hanle effect in the $90\deg$
scattering geometry of our observations are a reduction of the linear
polarization amplitude and a rotation of the direction of linear
polarization, with respect to the unmagnetized case.

The formation of the \HeI\,10830\,\AA\ triplet is sensitive to
both the Zeeman and Hanle effects.  We have taken advantage of this
fact to ``measure" the magnetic field in spicules.

\begin{figure}  
  \centering \includegraphics[width=2.6cm]{\figspath/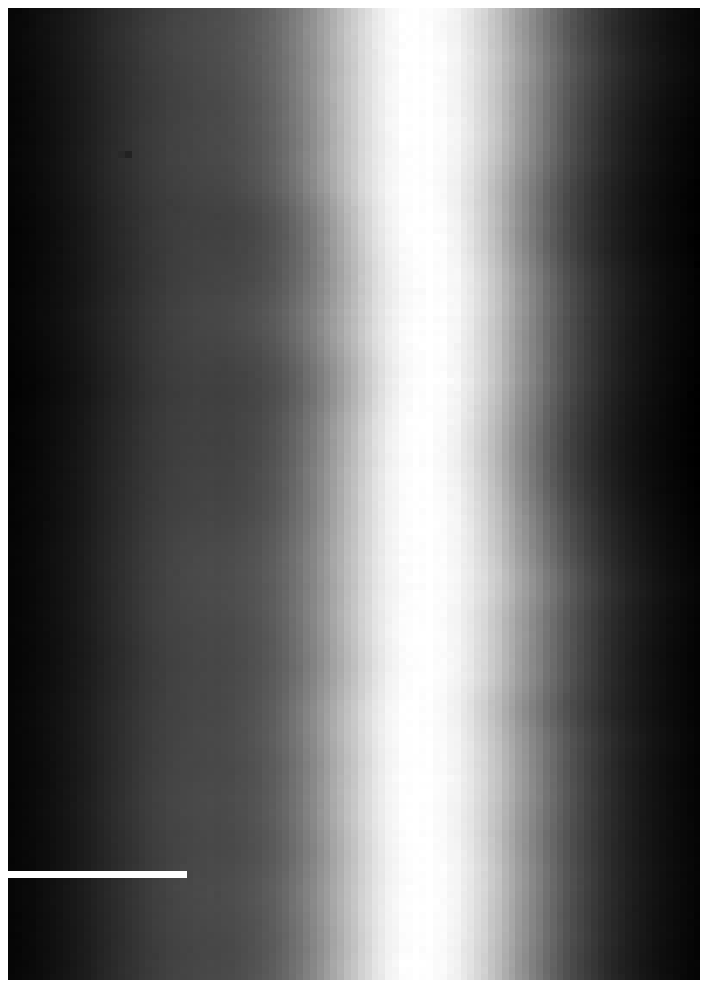}
  \includegraphics[width=2.6cm]{\figspath/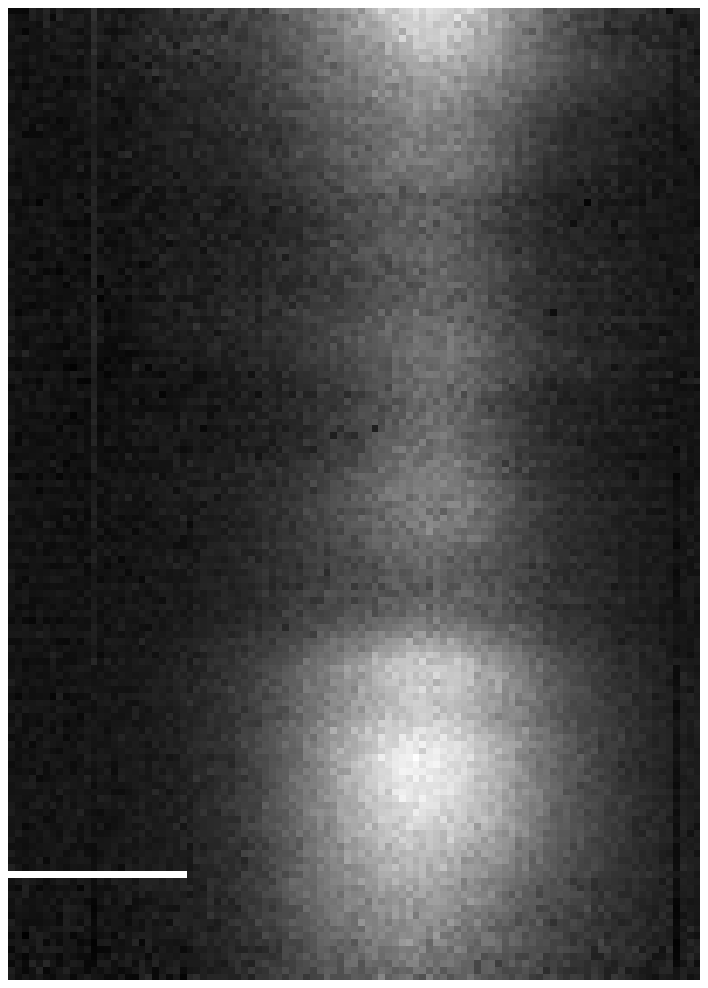}
  \includegraphics[width=2.6cm]{\figspath/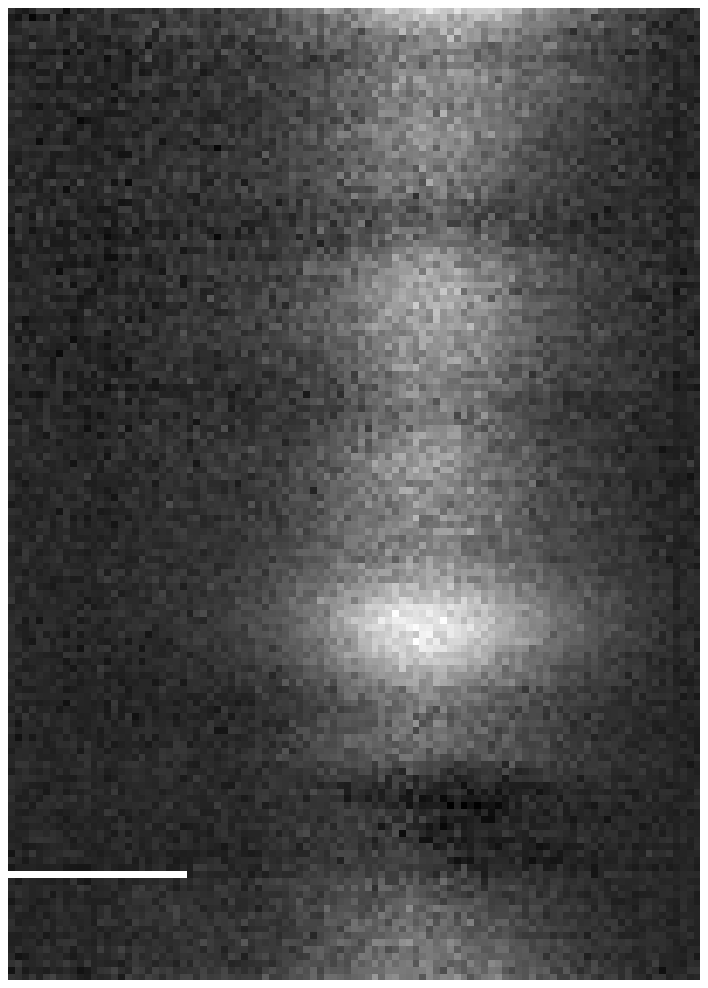}
  \includegraphics[width=2.6cm]{\figspath/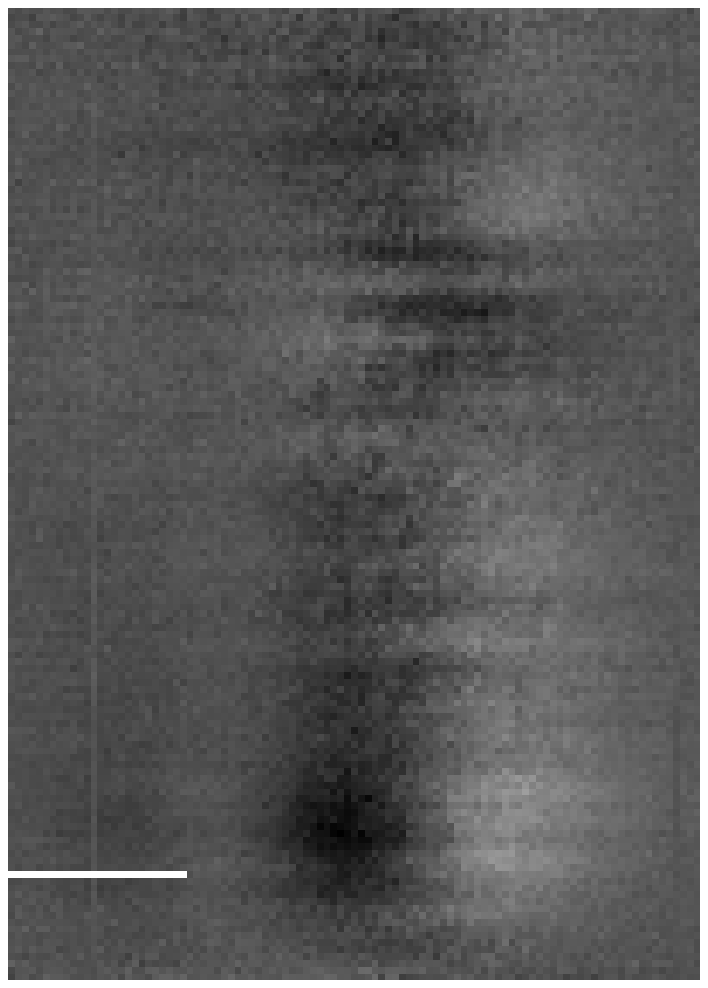}
  \caption[]{\label{centeno-fig:maps} From left to right, maps of
  Stokes $I$, $Q$, $U$ and $V$. The $x$-axis represents wavelength
  (increasing to the right), the $y$-axis the position along the slit
  (which is about 80\arcsec\ long).
\label{centenofig:stokesims} 
}
\end{figure}

\subsection{Detection of Zeeman-induced Stokes $V$}\label{centeno-stokesV}

The left-most panel of Fig.~\ref{centenofig:stokesims} shows
time-averaged intensity as a function of wavelength and position along
the slit.  The bright and fainter vertical strips correspond to the
red and blue components of the He multiplet, respectively.  Stokes $I$
provides physical and thermodynamical information: the damping, the
Doppler width, the optical depth and the macroscopic velocity of the
plasma.  Combining it with the information carried by Stokes $Q$ and
$U$ (second and third panels, respectively), one can infer the
magnetic field orientation. However, in the Hanle saturation regime
(which is above about 8~G for this multiplet), linear polarization is
barely responsive to the magnetic field strength, hindering its
determination.

\begin{figure}  
  \centering
  \includegraphics[width=9cm]{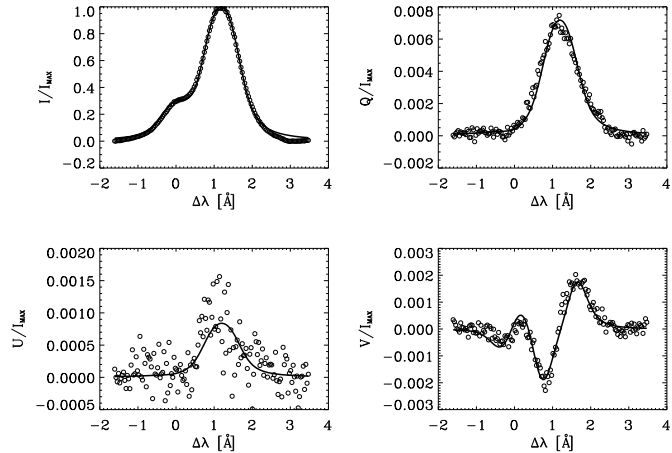}
  \caption[]{\label{centeno-hazelinv}
  Best fit produced by HAZEL for a set of Stokes profiles for one
  position along the slit that has a sizable Stokes-$V$ signal, with
  $B=36$ G, $\theta_B=38.6\deg$ and $\chi_B=-2.2\deg$. The reference
  direction for $Q$ is the tangential to the closest solar limb.
\label{centenofig:hazelfit}
}\end{figure}

One of the most striking findings in this
particular observation was the clear detection of a Zeeman-induced
Stokes-$V$ signature (right-most panel of
Fig.~\ref{centeno-fig:maps}). The circular polarization signal was, in
many cases, large enough that it allowed us to pin down magnetic field
strengths beyond the Hanle saturation value.  The antisymmetric Stokes
$V$ profile must be produced by a net line-of-sight (LOS) component of
the magnetic field, $B_{\rm LOS}$, that would be fully
resolved. However, cancellation effects due to the unresolved magnetic
structure in our spatio-temporal resolution element (or along the
line-of-sight), make the inferred $B_{\rm LOS}$ a lower limit for the
field strength. For the profiles in Fig.~\ref{centenofig:hazelfit}
(indicated by the horizontal lines in Fig.~\ref{centeno-fig:maps}),
the inferred value is $B_{\rm LOS} \approx 25$ G, provided by HAZEL.

How do we interpret this signal? In a picture where the spicules are
oriented arbitrarily along the line of sight, we would expect the $B_{\rm
LOS}$ to cancel, producing no net Zeeman Stokes V. However, these data
would seem to imply a preferred direction of the magnetic field.
Along the slit there are areas of strong and weak Stokes-$V$ signals,
suggesting that, in the latter cases, the magnetic field is 
conspiring to minimize (or, at least, reduce) the net LOS component.

\subsection{Inversions}

In order to determine the magnetic field strength and other physical
quantities from the observations, we inverted the full Stokes vector
for every position along the slit using the inversion code HAZEL (see
\cite{hazel}) to complete this task. HAZEL accounts for the physical
ingredients and mechanisms operating in the generation of polarized
light in this kind of observations: optical pumping, atomic level
polarization, and the Hanle and Zeeman effects. Radiative transfer is
computed in a constant-property slab that is permeated by a
deterministic magnetic field. The slab is located at height $h$
above the visible solar surface, and is illuminated by the CLV of the
photospheric continuum. The slab's optical thickness, $\tau$, accounts for
the integrated number of emitters and absorbers along the line of
sight, taking care of the collective effect of having several spicules
interposed along the path (although we cannot prescribe how many).

Fig.~\ref{centenofig:hazelfit} is an example in which our measurement
shows a sizable Stokes-$V$ profile, at the location shown in
Fig.~\ref{centeno-fig:maps}. The open circles represent the observed
profiles while the solid line shows the best fit (in a $\chi^2$ sense)
obtained from a HAZEL inversion.  The inferred magnetic field strength
is 36~G, with inclination 38.6\deg\ from the solar local
vertical, and azimuth $-2.2\deg$ with respect to the LOS.  The
magnetic field orientation is very well constrained by the observed
Stokes $Q$ and $U$ profiles. Except for the 180\deg\ and the
Van-Vleck ambiguities (see \cite{hazel} and references therein),
a good fit is only possible in a very narrow range of values.  However,
the field strength is only well determined when the Stokes-$V$ signal
is present.

We applied this inversion procedure to all the pixels along the slit
and both data-sets, deriving the magnetic field for all the spatial
positions at two heights above the visible limb.  From these inversion
we were able to trace the magnetic field vector and construct a
reliable picture of its behavior along the spicules. Variations of the
field strength and topology were detected which we will describe in
detail in forthcoming publications. Likewise, new observations
will be carried out in 2009 to complement this preliminary work.

\section{Conclusion}                   \label{centeno:conclusion}

We carried out spectropolarimetric measurements of quiet Sun spicules
in the \HeI\,10830\,\AA\ triplet, detecting clear Stokes $V$ signals
that allow us to infer magnetic field strengths beyond the Hanle
saturation regime.  Values as high as 40~G were found in localized
regions of the slit, which may correspond to organized network
spicules or perhaps a macro-spicule.

We determined the magnetic field vector of all the pixels along the
slit at two heights from the South limb, detecting spatial
variations in the magnetic field strength and orientation. We plan to
pursue this investigation further with new observations in the
10830\,\AA\ multiplet complemented with other useful data, such as
\Halpha\ or \CaII\ filtergrams and He\,D$_3$ spectropolarimetry.

\begin{acknowledgement}
The National Center for Atmospheric Research is sponsored by the
US National Science Foundation.  Financial support by the Spanish
Ministry of Science through project AYA2007-63881 and by the European
Commission via the SOLAIRE network (MTRN-CT-2006-035484) are
gratefully acknowledged.

\end{acknowledgement}

\begin{small}


\bibliographystyle{rr-assp}       

\begin{thebibliography}{10}
\expandafter\ifx\csname natexlab\endcsname\relax\def\natexlab#1{#1}\fi

\bibitem[{{Asensio Ramos} {et~al.}(2008){Asensio Ramos}, {Trujillo Bueno}, \&
  {Landi Degl'Innocenti}}]{hazel}
{Asensio Ramos}, A., {Trujillo Bueno}, J., {Landi Degl'Innocenti}, E. 2008,
  \apj, 683, 542

\bibitem[{{Beckers}(1972)}]{beckers-spicules}
{Beckers}, J.~M. 1972, \araa, 10, 73

\bibitem[{{De Pontieu} {et~al.}(2004){De Pontieu}, {Erd{\'e}lyi}, \&
  {James}}]{DePontieu}
{De Pontieu}, B., {Erd{\'e}lyi}, R., {James}, S.~P. 2004, \nat, 430, 536

\bibitem[{{De Pontieu} {et~al.}(2007){De Pontieu}, {McIntosh}, {Hansteen},
  {Carlsson}, {Schrijver}, {Tarbell}, {Title}, {Shine}, {Suematsu}, {Tsuneta},
  {Katsukawa}, {Ichimoto}, {Shimizu}, \& {Nagata}}]{depontieu2007}
{De Pontieu}, B., {McIntosh}, S., {Hansteen}, V.~H., {et~al.} 2007, \pasj, 59,
  655

\bibitem[{{L{\'o}pez Ariste} \& {Casini}(2005)}]{lopezariste-spicules}
{L{\'o}pez Ariste}, A. {Casini}, R. 2005, \aap, 436, 325

\bibitem[{{Mart{\'i}nez Pillet} {et~al.}(1999){Mart{\'i}nez Pillet},
  {Collados}, {S{\'a}nchez Almeida}, {Gonz{\'a}lez}, {Cruz-Lopez}, {Manescau},
  {Joven}, {Paez}, {Diaz}, {Feeney}, {S{\'a}nchez}, {Scharmer}, \&
  {Soltau}}]{martinezpillet-tip}
{Mart{\'i}nez Pillet}, V., {Collados}, M., {S{\'a}nchez Almeida}, J., {et~al.}
  1999, in High Resolution Solar Physics: Theory, Observations, and Techniques,
  eds. T.~R. {Rimmele}, K.~S. {Balasubramaniam}, \& R.~R. {Radick}, ASP Conf.
  Ser., 183, 264

\bibitem[{{Ramelli} {et~al.}(2006){Ramelli}, {Bianda}, {Merenda}, \& {Trujillo
  Bueno}}]{ramelli-spicules}
{Ramelli}, R., {Bianda}, M., {Merenda}, L., {Trujillo Bueno}, T. 2006, ASP
  Conf. Ser., 358, 448

\bibitem[{{Sterling}(2000)}]{sterling-spicules}
{Sterling}, A.~C. 2000, \solphys, 196, 79

\bibitem[{{Trujillo Bueno}(2005)}]{tb-quantumspectropol}
{Trujillo Bueno}, J. 2005, in The Dynamic Sun: Challenges for Theory and
  Observations, ESA Special Publication, 600

\bibitem[{{Trujillo Bueno} {et~al.}(2005){Trujillo Bueno}, {Merenda},
  {Centeno}, {Collados}, \& {Landi Degl'Innocenti}}]{tb-spicules}
{Trujillo Bueno}, J., {Merenda}, L., {Centeno}, R., {Collados}, M., {Landi
  Degl'Innocenti}, E. 2005, \apjl, 619, L191

\end{thebibliography}

\end{small}

\end{document}